\documentclass[
    ,final            % use final for the camera ready runs
  ]
  {aipproc}

\layoutstyle{8x11single}

\usepackage[usenames]{color}
\usepackage{comment}
\usepackage{amsmath}
\usepackage[utf8]{inputenc}

\def\beq{\begin{equation}}
\def\eeq{\end{equation}}
\def\bea{\begin{eqnarray}}
\def\eea{\end{eqnarray}}
\def\D0{D\O }
\def\ra{\rightarrow}

\newcommand*\xbar[1]{%
  \hbox{%
    \vbox{%
      \hrule height 0.5pt % The actual bar
      \kern0.5ex%         % Distance between bar and symbol
      \hbox{%
        \kern-0.1em%      % Shortening on the left side
        \ensuremath{#1}%
        \kern-0.1em%      % Shortening on the right side
      }%
    }%
  }%
}

\begin{document}

\title{Semileptonic and leptonic B decays, circa 2016}


\author{Giulia Ricciardi}{
  address={Dipartimento di Fisica E. Pancini, Universit\`a  di Napoli Federico II \\
Complesso Universitario di Monte Sant'Angelo, Via Cintia,
I-80126 Napoli, Italy\\
and \\
 INFN, Sezione di Napoli\\
Complesso Universitario di Monte Sant'Angelo, Via Cintia,
I-80126 Napoli, Italy}
}

\begin{abstract}
We summarize the status  of  semileptonic and leptonic $B$ decays, including $|V_{cb}|$ and $|V_{ub}|$ exclusive and inclusive determinations,
 decays to excited states of the charm meson spectrum  and decays into $\tau$ leptons.

\end{abstract}

\maketitle

%%%%%%%%%%%%%%%%%%%%%%%%%%%%%%%%%%%%%%%%%%%%
%% MAINMATTER
%%%%%%%%%%%%%%%%%%%%%%%%%%%%%%%%%%%%%%%%%%%%

\section{Introduction}

Today accuracy in measurements and theoretical calculations is indispensable to check the  Standard Model (SM) and explore the small region of parameters space  left to its extensions, at  our energies.
Semi-leptonic and leptonic $B$ decays are well suited to respond to such necessity.
The heavy mass of the $B$ meson allows to exploit simplifications  in the limit of infinite quark mass and to better separate low and high energy regimes. Past, present and future $B$ factories have provided and will provide  an unparalleled level of precision in branching ratios and related observables, and LHCb is following through.
One example is given by the determination of the  values of the CKM parameters  $|V_{cb}|$ and $|V_{ub}|$, which
strongly affects the identification of new physics \cite{Buras:2013ooa}.
%At present, $|V_{ud}|$ is the best known CKM parameter, with a relative uncertainty of the order $10^{-4}$,  all other %$|V_{ij}|$ CKM parameters are known at \% level.
% The values of
%$|V_{cd}|$
%from leptonic
%and semileptonic decays agree, while those for $|V_{cs}|$ are marginally compatible at the 1.1
%$\sigma$ level \cite{Rosner:2015wva}.
%$|V_{ub}|$ stands as having the last precise estimate, reaching in some determinations an uncertainty of 10\%
%
%are
%At present, the most precise values of  $|V_{cb}|$ and $|V_{ub}|$ come from  inclusive and exclusive  semileptonic decays.
%
 The   inclusive and exclusive semi-leptonic  searches  rely on
different theoretical calculations  and on
different experimental techniques which have, to a large extent, uncorrelated
statistical and systematic uncertainties. This independence makes
the comparison of $|V_{cb}|$ and $|V_{ub}|$ values from inclusive and exclusive decays an interesting test of our physical understanding\footnote{For recent reviews see e.g. R.  Kowalewski and T. Mannel in \cite{Olive:2016xmw}, Refs. \cite{Ricciardi:2014iga,  Ricciardi:2013cda, Ricciardi:2013jf, Ricciardi:2013xaa, Ricciardi:2012pf, Ricciardi:2012dj, Brambilla:2014jmp, Ricciardi:2016jjb} and references therein.}.
Extensive investigations have strongly reduced the uncertainties on $|V_{cb}|$ and $|V_{ub}|$ (amounting to about 5\% and 15\%, respectively, in the 1999 LEP determinations \cite{Calvi:1999si}), but a tension between inclusive and exclusive determination  remains to this day.

Here, we summarize   the status  of  semileptonic and leptonic $B$ decays,  mediated at lower parton level by tree diagrams, including
 decays to excited states of the charm meson spectrum  and decays into $\tau$ leptons.

\section{Heavy-to-heavy semileptonic decays}

\subsection{Exclusive decays into light leptons}
\label{subsectionExclusive decays}

For negligible lepton masses ($l=e, \mu)$,
the  differential ratios for the semi-leptonic CKM favoured decays $B \to D^{(\ast)} l  \nu$
can be written as
\begin{eqnarray}
&\frac{d\Gamma}{d \omega}(B \rightarrow D^\ast\,l {\nu})&
=  \frac{G_F^2}{48 \pi^3}  (m_B-m_{D^\ast})^2 m_{D^\ast}^3 \chi (\omega)  (\omega^2-1)^{\frac{1}{2}}
 |V_{cb}|^2  |\eta_{EW}|^2 |{\cal F}(\omega)|^2
 \nonumber \\
 &\frac{d\Gamma}{d \omega} (B \rightarrow D\,l {\nu})&  =
 \frac{G_F^2}{48 \pi^3}\,   (m_B+m_D)^2  m_D^3 \,
(\omega^2-1)^{\frac{3}{2}}\,
 |V_{cb}|^2  |\eta_{EW}|^2 | {\cal G}(\omega)|^2
 \label{diffrat}
\end{eqnarray}
in terms of a single form factor ${\cal F}(\omega)$ and  ${\cal G}(\omega)$, for $B \to D^{\ast} l  \nu$ and $B \to D l  \nu$, respectively.
 In Eq. (\ref{diffrat}),  the differential cross sections are proportional to $|V_{cb}|^2$,
  $\eta_{EW}$ is a structure-independent one-loop  electroweak enhancement factor and $\chi (\omega)$  is a
 phase space factor  which reads
 \beq \chi (\omega)= (\omega+1)^2 \left( 1 + \frac{ 4 \omega}{\omega + 1} \frac{ m_B^2 - 2 \omega m_B m_D^\ast + m_{D^\ast}^2}{(m_B-m_{D^\ast})^2} \right)\eeq
The  recoil parameter $\omega = p_B \cdot p_{D^{(\ast)}}/m_B \, m_{D^{(\ast)}}$ corresponds to the energy transferred to the leptonic pair. In the
heavy quark limit
both form factors are related to a single Isgur-Wise
function,  ${\cal F(\omega) }= {\cal G(\omega) } = {\cal  \xi (\omega) }  $, which is
normalized to unity at zero recoil,  that is  ${\cal \xi (\omega}=1) =1 $.
There are perturbative and
nonperturbative corrections  to this prediction, the latter
% suppressed by powers of $\alpha_s(m)/\pi$
 suppressed by powers of $\Lambda_{QCD}/m$, where $m= m_c$ and $m_b$.

At zero recoil, the heavy quark symmetries provide the structure of the symmetry breaking non-perturbative corrections, which start at order $1/m^2 $ and  $ 1/m$ for the  ${\cal F(\omega}=1)$ and ${\cal G(\omega}=1)$ form factors, respectively.
The downside of zero-recoil
analyses  is that, since  decay rates vanish at zero-recoil,
 one  needs to extrapolate  the experimental points to zero recoil, using a parameterization of the momentum
dependence. There are several parameterizations for the momentum dependence of the form factors, that
generally fall under two categories i)
  based on a simple pole, as
 the BZ (Ball-Zwicky) \cite{Ball:2001fp}
at 4 parameters, or the BK
(Becirevic-Kaidalov) at 3 parameters \cite{Becirevic:1999kt} ii)
based on a series
expansion, where
 $\omega$ is mapped onto a complex variable $z(\omega)$ via the conformal transformation
$
z= \frac{\sqrt{\omega+1}-\sqrt{2}}{\sqrt{\omega+1}+\sqrt{2}}
$. The
expansion in $z$ converges rapidly in the kinematical
region of heavy hadron decays.
 Common examples are the
 CLN
(Caprini-Lellouch-Neubert) \cite{Caprini:1997mu},
 the BGL
(Boyd-Grinstein-Lebed)  \cite{Boyd:1994tt} and the BCL (Bourrely-Caprini-Lellouch) \cite{Bourrely:2008za}  parameterizations.

Several computations of form factors have been performed on lattice.
 The difficulties related to heavy fermions  can be n\"aively summarized by observing that direct simulation of high mass such $m a \geq 1$, where $a$ represent a lattice spacing, gives discretization errors out of control.   As of today
$ m_b \sim 1/a
$ and  no direct simulation is possible. The main
 way out is  the usage of
effective theories, as Heavy Quark Effective Theory (HQET)
\cite{Isgur:1989vq} and Non-Relativistic QCD (NRQCD) \cite{Caswell:1985ui}. In broad terms, they eliminate high degrees of freedom  relying  on a systematic expansion in $\Lambda_{QCD}/m_b$.
The downside is the introduction of new sources of errors (matching of HQET to QCD, renormalization, control of extrapolation,etc.) to take care of.

Another common approach to non-perturbative calculations of form factors are  QCD sum rules.
The sum rules are  based on the general idea of calculating a relevant quark-current correlation function and relating it to the hadronic parameters of interest via a dispersion relation. They have reached  wide application for calculation of exclusive  amplitudes and form factors in the form of
light cone sum rules (LCSR), employing
 light-cone operator product expansion
(OPE) of the relevant  correlation functions. Uncertainties may originate from the truncation of the expansions, the input parameter uncertainties, and
the assumption of quark-hadron duality.

Let us first consider the $B \to D^{\ast} l  \nu$ channel, which  is less suppressed in the phase space and whose branching fractions are  more precise (even twice) in the majority of experimental measurements.
On lattice,  the  progress on the $B\to D^\ast$ form factor is slower, since
this channel poses greater technical complications than
the $B\to D$, due to the fact that the $D^\ast$ is unstable.
The FNAL/MILC  collaboration has  performed
the  non perturbative determination  of the form factor ${\cal F}(1)$
in the lattice unquenched $N_f= 2+1$  approximation \cite{Bernard:2008dn, Bailey:2014tva}. The FNAL/MILC  collaboration
uses FNAL $b$-quark and asqtad $u$, $d$, $s$ valence quarks. Their latest result at zero recoil exploits
the full suite of MILC (2+1)-flavor asqtad ensembles for sea quarks and  gives the following estimate\cite{Bailey:2014tva}.
\beq  {\cal F}(1)
=0.906\pm 0.004 \pm  0.012  \label{VcbexpF2}  \eeq
The first error is statistical and the second one  is the sum in quadrature of all systematic errors. Using the previous form factor
and the 2012 HFAG average \cite{Amhis:2012bh}
 the following estimate for $|V_{cb}| $ has been given \cite{Bailey:2014tva}
\beq |V_{cb}| = (39.04 \pm 0.49_{\mathrm{exp}} \pm 0.53_{\mathrm{latt}} \pm 0.19_{\mathrm{QED}}) \, \mathrm{x} \, 10^{-3} \label{ll1} \eeq
which is  reported in  Table \ref{phidectab2}. The central value is not very different from the central value of the 2009 determination from the same Collaboration \cite{Bernard:2008dn}, but errors are considerably reduced.
The
analysis strategies are similar, but the lattice-QCD data set is
much  more  extensive,  with  higher  statistics  on  all  ensembles, smaller  lattice spacings (as small as 0.045 fm) and light-to-strange-quark mass ratios  can be as low as 1/20.
The lattice QCD theoretical error is now commensurate with the experimental error, they contribute respectively for about 1.4\% and 1.3\%, while  the QED error contributes for about  0.5\%. Largest QCD errors come from discretization and
are estimated taking the difference between
HQET description of lattice gauge theory and QCD.
The discretization error could be in principle reduced  by going to finer lattice spacings
or by using a more improved Fermilab action.
Subleading errors appearing  at the 0.4-0.6\% level are
nontrivial to improve.  Reducing the error from the QED Coulomb correction would require
a detailed study of electromagnetic effects within HQET, and reducing the QCD matching
error would require a two-loop lattice perturbation theory calculation or nonperturbative
matching \cite{Bailey:2014tva}.

Other, preliminary, values for the $B \to D^{\ast}$ form factor at zero recoil, in agreement with the value reported  in \eqref{VcbexpF2}, were also obtained at $N_f=2$, by using charmed
quarks having a realistic finite mass and two
ensembles of gauge configurations produced by the European
Twisted Mass Collaboration (ETM) \cite{Atoui:2013sca}.
%At a variance with the approach used by the FNAL/MILC collaboration,  in Ref. \cite{Atoui:2013sca} form factors and %then the branching ratios are determined

The 2016 FLAG  $N_{f}=2+1$ $|V_{cb}|$  average value yields \cite{Aoki:2016frl}
\beq |V_{cb}| = (39.27 \pm 0.49_{\mathrm{exp}} \pm 0.56_{\mathrm{latt}}) \times\, 10^{-3} \label{ll1flag} \eeq
This average, reported in Table \ref{phidectab2}, employs  the latest HFAG experimental average  \cite{Amhis:2014hma} ${\cal F}(1) \eta_{\mathrm{EW}} |V_{cb}|= (35.81 \pm 0.45) \times\, 10^{-3}$, the value $\eta_{\mathrm{EW}} = 1.00662$, and the value  ${\cal F}(1) $ given in Eq.
\eqref{VcbexpF2}.

At the current level of precision, it would be important to extend
form factor unquenched  calculations for   $B \to D^{\ast}$ semileptonic decays
  to nonzero recoil, in order to reduce the uncertainty due to the extrapolation to $\omega=1$.
Indeed, at finite momentum transfer,  only old  quenched lattice results are  available \cite{deDivitiis:2008df} which,
 combined with 2008 BaBar data \cite{Aubert:2007rs},  give  $ |V_{cb}| = 37.4 \pm 0.5_{\mathrm{exp}} \pm 0.8_{\mathrm{th}}$.

Older  form factor  estimates are available via zero recoil sum rules, giving
 \cite{Gambino:2010bp, Gambino:2012rd}
\beq {\cal F}(1) = 0.86 \pm 0.02 \label{gmu} \eeq
in good agreement with the lattice value in Eq. \eqref{VcbexpF2}, but  slightly lower in the central value. That implies  a relatively higher value of $|V_{cb}|$, that is
\beq  |V_{cb}| = (41.6\pm 0.6_{\mathrm{exp}}\pm 1.9_{\mathrm{th}})  \, \mathrm{x} \, 10^{-3}
\label{wee}
\eeq
where the HFAG averages \cite{Amhis:2012bh} have been used. The theoretical error  is more than twice the error in the lattice determination \eqref{ll1}.

For $ B \rightarrow D \, l \, \nu$ decay,
unquenched lattice-QCD calculation of the hadronic form factors
at  nonzero  recoil have  become available in 2015,  due to the FNAL/MILC collaboration \cite{Lattice:2015rga}.
%\footnote{Preliminary results had been presented two years earlier \cite{Qiu:2013ofa}, giving the value
%$ |V_{cb}| =(38.5 \pm 1.9_{\mathrm{exp+QCD}} \pm 0.2_{\mathrm{QED}})   \, \mathrm{x} \, 10^{-3} $.
%The first error combines statistical and
%systematic errors from both experiment and theory. The second error reflects the uncertainty in
%the Coulomb correction.
%The new analysis  includes a more sophisticated treatment of the matching
%factors as well as more refined estimates for the renormalization and heavy-quark discretization %errors.}.
Prior results at non-zero recoil
%the  value for the form factor
 were only available   in the quenched approximation \cite{deDivitiis:2007ui, deDivitiis:2007uk} \footnote{By using  the step scaling approach  and
the 2009 data from BaBar Collaboration,  for $ B \rightarrow D \, l \, \nu$ decays \cite{Aubert:2009ac}, the value
  $|V_{cb}| = 37.4 \pm 0.5_{\mathrm{exp}} \pm 0.8_{\mathrm{th}} $ was obtained.
The errors are  statistical, systematic and due to the theoretical uncertainty in the form factor $ {\cal G}$, respectively.}.

%
%The error
%could be improved   by repeating the
%analysis with a world average of experimental form factors,
%and/or by ameliorating the  understanding of the experimental systematic error at large
%$\omega$ due to the vanishing phase space.
%To quantify the improvement due to working at nonzero recoil,
% $|V_{cb}| $  is also extracted by extrapolating the experimental data to zero recoil and comparing
%with the theoretical form factor at that point.
%The result gives a value found consistent with the  nonzero recoil determination,   within the (expected) larger error %\cite{Qiu:2013ofa}.
%
In \cite{Lattice:2015rga}   the FNAL/MILCcollaboration has calculated the form factors
for a range of recoil momenta and parameterized their dependence on momentum transfer
using  the BGL z-expansion, determining
$ |V_{cb}| $
from the relative normalization over the entire range of recoil momenta.
Their estimate gives \cite{Lattice:2015rga}
\beq
 |V_{cb}| =(39.6 \pm 1.7_{\mathrm{exp+QCD}} \pm 0.2_{\mathrm{QED}})   \, \mathrm{x} \, 10^{-3}
\eeq
%which has smaller experimental plus QCD uncertainty, but the same  QED one than their previous result.
The   average value is almost the same than the  one inferred from
$ B \rightarrow D^\ast \, l \, \nu$ decay by the same collaboration, see Eq. \eqref{ll1} and  Table \ref{phidectab2}.

% Heavy-quark discretization errors are the largest source
%of uncertainty   on   $|V_{cb}| $ determinations  by the FNAL/MILC collaboration using  both exclusive $ B \rightarrow %D^{\ast} \, l \, \nu$ and $ B \rightarrow D \, l \, \nu$ decays. Work is  in progress  to reduce them by improving the %Fermilab action to third order in HQET \cite{Jang:2013yqa}.

Two months later,  new results on
$ B \rightarrow D\, l \, \nu$
form factors at non-zero recoil were announced by the HPQCD Collaboration \cite{Na:2015kha}. Their results are based    on the non-relativistic QCD (NRQCD) action
for  bottom  and  the  Highly  Improved  Staggered  Quark
(HISQ) action for charm quarks, together with $N_f=2+1$ MILC gauge configuration.
A  joint  fit to  lattice  and 2009  BaBar  experimental  data  \cite{Aubert:2009ac} allows the
extraction of the CKM matrix element  $|V_{cb}| $, which reads
\beq
|V_{cb}|= (40.2 \pm 1.7_{\mathrm{latt+stat}} \pm 1.3_{\mathrm{syst}})   \, \mathrm{x} \, 10^{-3}
\eeq
The  first  error  consists  of  the  lattice  simulation  errors  and  the  experimental  statistical  error  and
the  second  error  is  the  experimental  systematic  error.
The dominant error  is the discretization error, followed by higher order current matching uncertainties.  The former error can be reduced by adding
simulation  data  from  further  ensembles  with  finer  lattice spacings.
% Simulations are under their way also to improve
%matching errors by combining different simulations.

%
%On the non-lattice front, the ''BPS" limit is
%the limit
%where the parameters related to kinetic energy
%and
%the chromomagnetic moment are equal in the heavy quark expansion \cite{Uraltsev:2003ye}.
%%
%Using this limit, the Particle Data Group finds the form factor  \cite{Beringer:1900zz}
 %\beq {\cal G}(1) =1.04 \pm  0.02 \eeq
%and
%the related
%\beq |V_{cb}| = (40.6 \pm 1.5_{\mathrm{exp}} \pm 0.8_{\mathrm{th}}) \, \mathrm{x} \, 10^{-3} \eeq

The same year, in order to interpret
the $\Delta \Gamma/\Delta \omega$ distribution, the Belle collaboration \cite{Glattauer:2015teq} has  performed a fit to the CLN parameterization, which has two free parameters, the form factor at zero recoil ${\cal G}(1)$ and the linear slope $\rho^2$.
The fit has been used to determine $\eta_{EW} {\cal G}(1) |V_{cb}|$, that, divided by the form-factor normalization ${\cal G}(1)$ found by the FNAL/MILC Collaboration
\cite{Lattice:2015rga}, gives
$
\eta_{EW}  |V_{cb}|=(40.12 \pm 1.34) \times 10^{-3}
$  \cite{Glattauer:2015teq}.
Assuming $\eta_{EW}   \simeq 1.0066 $, it  translates into  \cite{Glattauer:2015teq}\beq
 |V_{cb}|=(39.86 \pm 1.33) \times 10^{-3}\eeq The Belle Collaboration also obtain
a slightly more precise
result (2.8\% vs.  3.3\%)  by
 exploiting  lattice data at non-zero recoil and  performing   a combined fit to the BGL form factor. It yields
$
\eta_{EW}  |V_{cb}|=(41.10 \pm 1.14) \times 10^{-3}
$
which translates into \cite{Glattauer:2015teq} \beq
 |V_{cb}|=(40.83 \pm 1.13) \times 10^{-3}
\eeq
assuming once again $\eta_{EW}   \simeq 1.0066 $.

A very recent $ |V_{cb}|$ determination is in agreement with previous results, being \cite{Bigi:2016mdz} \beq
 |V_{cb}|=(40.49 \pm 0.97) \times 10^{-3} \eeq
It makes use of latest lattice \cite{Lattice:2015rga, Na:2015kha}, Belle \cite{Glattauer:2015teq} and Babar \cite{Aubert:2009ac} results, and consider the BGL, CLN, and BCL parameterizations, checking that they all yield consistent
results.

Semileptonic $B_s$ decays can also  probe CKM matrix elements. Moreover,
 semileptonic $B^0_s$
decays are also used as a normalization mode for various
searches for new physics at hadron colliders and at Belle-II.
On lattice, the valence strange quark needs less of a chiral extrapolation
and  is  better accessible in numerical simulations with respect to
 the physical $u(=d)$-quark.
Zero-recoil form factors for
 $ B_s \rightarrow D_s \, l \, \nu$ decays have now been computed at $N_f = 2$, for the first time, using the ETM (European twisted mass) approach \cite{Atoui:2013zza}.
They employ
 lattice spacings within the range $a \simeq  0.054$ to 0.098 fm, using the
maximally twisted Wilson quark action and obtain the result  $ {\cal G}(1)^{B_s \rightarrow D_s}= 1.052 \pm 0.046$,
which has an uncertainty of 4\%.

\subsection{Exclusive decays into heavy leptons}

The   $ B \to D^{(\ast)} \tau  \nu_\tau$ decays are more difficult to measure,
since  decays into the heaviest $\tau$ lepton are phase space suppressed and there are
 multiple neutrinos in the final state, following the $\tau$ decay, which
 stand in the way of the reconstruction of the invariant mass of $B$ meson.
% At the $B$ factories, a major constraint exploited is the fact that $B$ mesons are produced from the %process $e^+e^- \to \Upsilon (4 S)\to B \bar B$.

The ratio of branching fractions
\beq
{\cal{R}}(D^{(\ast)}) \equiv  \frac{{\cal{B}}( B \to D^{(\ast)} \tau \nu_\tau)}{{\cal{B}}( B \to D^{(\ast)} l  \nu_l)} \qquad \qquad (l=e, \mu)
\label{ratiotau0}
\eeq
is typically used instead of the absolute branching fraction
of $ B \to D^{(\ast)} \tau  \nu_\tau$ decays, to reduce several systematic uncertainties
such as those on the experimental efficiency and  on the form factors, and to eliminate the dependence on the value of  $|V_{cb}|$. In 2012, FNAL/MILC has published the first unquenched
lattice determination of ${\cal{R}}(D)_{SM}$ \cite{Bailey:2012jg}, whose update in 2015 \cite{Lattice:2015rga} is in excellent agreement with
the results presented in 2016 by the HPQCD Collaboration \cite{Na:2015kha}.
A SM phenomenological prediction \cite{Fajfer:2012vx} of ${\cal{R}}(D^{\ast})$ is currently available, but no lattice-based computations.
Summarizing, the most recently determinations are
 \bea
{\mathcal{R}}(D)_{SM} &=& 0.299 \pm 0.011 \quad \quad   \text{ FNAL/MILC \,\cite{Lattice:2015rga}}  \nonumber  \\
 {\mathcal{R}}(D)_{SM} &=& 0.300\pm 0.008 \quad \quad \text{HPQCD \, \cite{Na:2015kha} } \nonumber \\
  {\mathcal{R}}(D)_{SM} &=& 0.299 \pm 0.003 \quad \quad \text{ \cite{Bigi:2016mdz} } \nonumber \\
{\mathcal{R}}(D^\ast)_{SM}&=& 0.252\pm 0.003 \quad \quad \text{ \cite{Fajfer:2012vx}}
\eea
%is found
%in  a  (2+1)-flavor lattice QCD calculation
Older $ {\mathcal{R}}(D)_{SM}$ determinations \cite{Kamenik:2008tj, Becirevic:2012jf} are in agreement as well.

 Exclusive semi-tauonic $B$ decays were
first observed by the Belle Collaboration in 2007 \cite{Matyja:2007kt}. Subsequent
analysis by Babar and Belle \cite{Aubert:2007dsa, Bozek:2010xy,Huschle:2015rga},
 performed using an hadronic or
an inclusive tagging method, measured
 branching fractions above--yet consistent with--the SM predictions. In 2012-2013
Babar
 has measured
${\cal{R}}(D^{(\ast)})$ by using  its full data sample \cite{Lees:2012xj, Lees:2013uzd},
and reported a significant excess over the SM expectation, confirmed  in 2015
by LHCb \cite{Aaij:2015yra} and in 2016 by Belle \cite{Abdesselam:2016cgx}, the latter performing
the first
measurement of ${\cal{R}}(D^{(\ast)})$ at the $B$ factories  using the semileptonic tagging
method.
By averaging the  recent measurements  \cite{Huschle:2015rga,Lees:2012xj, Lees:2013uzd,Aaij:2015yra, Abdesselam:2016cgx}, the HFAG Collaboration has found \cite{HFAG2016}
\bea
{\cal{R}}(D)  &=& 0.397 \pm 0.040 \pm 0.028 \nonumber \\
{\cal{R}}(D^{\ast}) &=& 0.316 \pm 0.016 \pm 0.010
\label{ratiotau}
\eea
where the first uncertainty is statistical and the second is
systematic. ${\cal{R}}(D^{\ast})$ and ${\cal{R}}(D)$ exceed the SM
predictions in Refs \cite{Fajfer:2012vx} and \cite{Na:2015kha} by 3.3$\sigma$ and 1.9$\sigma$, respectively.
The combined analysis of ${\cal{R}}(D^{\ast})$ and ${\cal{R}}(D)$, taking
into account measurement correlations, finds that the deviation
is 4$\sigma$ from the SM prediction.
The first measurement of ${\cal{R}}(D^{\ast})$
using the semileptonic tagging method has been reported this year by the Belle Collaboration, giving
\beq
{\cal{R}}(D^{\ast}) = 0.302\pm 0.030\pm 0.011
\label{ratiotauBlle2016}
\eeq
which is within 1.6 $\sigma$ of the SM
expectation, where the standard
 deviation $\sigma$
includes systematic uncertainties.

Very recently, the Belle collaboration has presented
the  first measurement of the $\tau$
lepton polarization in the decay $\bar B \to D^\ast \tau^- \bar \nu$ as well as a new measurement of
 in the hadronic
$\tau$
decay modes
which is statistically independent of the previous Belle
measurements,  with  a  different  background  composition \cite{Abdesselam:2016xqt}.
The  preliminary results are  consistent with the theoretical predictions of the SM
 within 0.6$\sigma$ standard deviations, in particular \cite{Abdesselam:2016xqt}
$
{\cal{R}}(D^{\ast}) = 0.276\pm 0.034^{+0.029}_{-0.026}
$.

%In the case of ${\cal{R}}(D)$, other determinations are available, which are
% consistent with the HFAG analyses  \cite{Kamenik:2008tj}.

%
%The latest data from BaBar are not compatible with a
%charged Higgs boson in the type II two-Higgs-doublet model
%and with large portions of the more general type III two-Higgs-doublet model \cite{Lees:2013uzd}.
%
%At present, semi-leptonic  $b \to \tau$  decays
%do  not  contribute to the determination
%of $|V_{cb}|$,  but are studied because of their NP sensitivity.
%
The alleged   breaking of lepton-flavour universality suggested by most of the data is quite large,  and several theoretical models have been tested against the experimental results:
minimal flavor violating models,
 right-right vector and right-left scalar quark currents,  leptoquarks, quark and lepton
compositeness models \cite{Fajfer:2012jt, Sakaki:2013bfa,  Bauer:2015boy, Becirevic:2016yqi}, modified couplings  \cite{Abada:2013aba, Datta:2012qk},  additional tensor operators  \cite{Biancofiore:2013ki}, charged scalar contributions \cite{Dorsner:2013tla},
  effective Lagrangians
\cite{Fajfer:2012vx, Datta:2012qk}, new sources of CP violation \cite{Hagiwara:2014tsa}, quantum effects \cite{Feruglio:2016gvd}, type III two-Higgs-doublet  \cite{Crivellin:2012ye}, R-parity violating Susy  \cite{Deshpand:2016cpw} and lepton-flavout non universal $SU(2)\times SU(2)\times U(1)$ models \cite{Boucenna:2016qad},  just to quote some.
A welcome feature of measurements in the $\tau$ sector is the capacity of putting  stringent limits on new physics models (see e.g. \cite{Celis:2012dk, Faroughy:2016osc, Becirevic:2016hea, Bordone:2016tex}).
%The  A2HDM does not seem able to accomodate present data on  $ {\cal{R}}(D) $ %\cite{Celis:2014pza,Celis:2012dk }.

At Belle II, with more data, there will be a better understanding of
backgrounds tails under the signal. At 5 ab$^{-1}$ the expected uncertainty is of 3\% for ${\cal{R}}(D^{\ast})$  and 5\% for   ${\cal{R}}(D)$. Data from Belle II may in principle be used for  the inclusive $ B \ra X_c \tau \nu$  decays, where
 predictions for the
 dilepton invariant mass and the $\tau$ energy
distributions  already exist \cite{Ligeti:2014kia}.

In the context of leptonic non-universality, it is worth mentioning the violation observed at LHCb in the $B \to K l^+ l^-$ channel \cite{Aaij:2014ora}.
Across the dilepton
invariant-mass-squared range 1 GeV$^2<m_{ll}^2 < 6$ GeV$^2$, the ratio $R(K)= \mathrm{Br}[B \to K \mu^+ \mu^-]/\mathrm{Br}[B \to K e^+ e^-]=0.745^{+0.090}_{-0.074}\pm 0.036$  disagrees with the theoretically clean SM prediction $R(K)=1.0003 \pm 0.0001$ by 2.6$\sigma$
\cite{Bobeth:2007dw}.

\subsection{$B$-Mesons Decays to Excited $D$-Meson States}

The increased interest in
semi-leptonic $B$ decays to excited states of the
charm meson spectrum  derives  by the fact that they
contribute
as a background to the direct decay $ B \to D^{(\ast )} l  \nu$ at the B factories, and, as a consequence, as
 a source of systematic error in the $|V_{cb}|$ measurements.

The spectrum of mesons consisting of a charm and an
up or a down anti-quark (open charm mesons) is poorly known.
%In the non-relativistic constituent quark model,
%it  can be classified according to  the radial quantum
%number and to the eigenvalue $L$ of the relative angular momentum
%between  the c-quark and the light degrees of freedom,
%
A QCD framework for their analysis can be set
up by using  HQET.
In the limit  of infinite heavy quark mass,   the spin  of the heavy quark $\vec{s_h}$
 is conserved
and  decouples from the total angular momentum of the light
degrees of freedom $\vec{j_l}$, which  becomes  a conserved quantity as well.
The
separate conservation in strong interaction processes of $\vec{s_h}$ and $\vec{j_l}$  permits a classification of heavy mesons of given radial (principal)  quantum
number according to the
value of $\vec{j_l}$.
Mesons can be collected in doublets: the two states in each doublet (spin partners)
have total angular momentum $\vec{J} = {\vec{j_l}} + 1/2 \, \hat{s_h}$ and parity
$P = (-1)^{L+1}$,  since ${\vec{j_l}} \equiv   {\vec{L}+{\vec{s_l}}}$, where $\vec{L}$ is the orbital angular momentum  and $\vec{s_l}$  the spin of the light degrees of freedom.
Within each doublet
the two states are degenerate in the limit of infinite heavy quark mass,
% and, due to flavour symmetry, the properties of
%the states in a doublet can be related to those of the corresponding states differing for the heavy
%quark flavour ($c \leftrightarrow b$). Corrections can be systematically included by considering next-to-leading terms in
%an expansion in $\/m_c$.

The low-mass spectrum includes the ground states, with principal (radial) quantum number $n=1$ and  $L = 0$ (1S, in the spectroscopic notation), which implies ${j_l}^P={\frac{1}{2}}^-$. The ground state doublet consists of two
states with $J^P = (0^-, 1^-)$,
that is $D$ and $D^\star$ mesons.

%The orbital excitations with angular momentum $L = 1, 2$
%(1P, 1D), and the first radial excitations (2S).

When $L=1$, there are four states ($1P$ states), which are generically referred to as $D^{\star \star}$ \footnote{Sometimes in literature this term is extended to include all particles in the low-mass spectrum
except the ground states.}. The doublet having ${j_l}^P={\frac{1}{2}}^+$ is named ($D_0^\ast, D_1^\prime$) and corresponds to $J^P=(0^+, 1^+)$. The doublet having ${j_l}^P={\frac{3}{2}}^+$ is named ($D_1, D_2^\ast$) and corresponds to $J^P=(1^+, 2^+)$.
These states are generally identified with $D_0^\ast(2400)$, $D_1^\prime(2430)$,
%(or $D_1^\star(2430)$),
$D_1(2420)$ and $D_2^\ast(2460)$ \footnote{The naming convention followed is to use $D^\ast(mass)$ to denote the states having $P=(-1)^J$, that is $J^P=0^+,1^-,2^+, \dots$ (natural spin-parity)and with $D(mass)$ all the others (unnatural spin-parity); the prime is used to distinguish between the two doublets.}.
%
%The orbital excitations with angular momentum $L = 1$ can have $j_l=3/2$ or $j_l=1/2$. The lowest masse state is $D_0^\ast(2400)$, one of the two states obtained componing
% $j_l=3/2$  with the heavy quark spin, that is the one with  $J^P=0^+$, the other one being  $D_1^\star(2430)$ (also said $D_1^\prime$), with  $J^P=1^+$. When $j_l=3/2$ is composed with the heavy quark spin, other two states,  $D_1(2420)$ and $D_2^\ast(2460)$, are originated, with $J^P=1^+$ and $J^P=2^+$, respectively.
 %The four states with L = 1 are generically denoted as $D^{\star \star}$ \footnote{Sometimes in literature this term is extended to include all particles in the low-mass spectrum
%except the ground states.}
%%%%%%%%%%%%%%%%%%%%%%%
$D_1(2420)$ and $D^\ast_2(2460)$
 have relatively narrow widths, about 20-30 MeV, and have been observed and studied  by a number of experiments
since the nineties (see Ref. \cite{Aubert:2009wg} and references therein).
The  other
two  states,  $D^\ast_0(2400)$, $D_1^\prime(2430)$,  are more difficult to detect due to the large width, about 200-400 MeV, and their observation has started more recently
\cite{Abe:2003zm, Abazov:2005ga, Abdallah:2005cx,Aubert:2008ea,Liventsev:2007rb}.
%
%The experimental results are at variance with theoretical expectations, since the states with large width should correspond to $j_l=1/2^+$ states, which decay as $ D_{0,1}^\ast \rightarrow D^{(\ast)} \pi $ through $S$ waves by conservation of parity and angular momentum. Similarly, the states with small width should correspond to $j_l=3/2^+$ states, since  $ D_2^\ast \rightarrow D^{(\ast)} \pi $  and $ D_1 \rightarrow D^{\ast} \pi $  decay  through $D$ waves. To be precise, the
%$ D_1 \rightarrow D^{\ast} \pi $ decays may occur a priori through $D$ and $S$ waves, but the latter are disfavored by heavy quark symmetry.

The spectroscopic identification for heavier states is
less clear.
In 2010 BaBar has observed,  for the first time, candidates for the radial excitations of the $D^0$, $D^{\ast 0}$ and $D^{\ast +}$, as well as the $L=2$ excited states of the $D^0$ and $D^+$ \cite{delAmoSanchez:2010vq}.
Resonances in the $2.4$-$2.8$  ${\mathrm{GeV/c}}^2$ region  of hadronic masses have also  been identified at LHCb
\cite{Aaij:2013sza, Aaij:2015vea, Aaij:2015sqa, Aaij:2016fma}.

%%%%%%%%%%%%%%%%%%%%%%%%%%%%%%%

%

The not completely clear experimental situation is mirrored by two theoretical puzzles.
Most   calculations, using sum rules \cite{LeYaouanc:1996bd,Uraltsev:2000ce}, quark models \cite{Morenas:1996yq, Morenas:1997nk,  Ebert:1998km, Ebert:1999ga},  OPE \cite{Leibovich:1997em, Bigi:2007qp} (but not   constituent quark models \cite{Segovia:2011dg}),
indicate that  the narrow width states dominate over
the broad  $D^{\ast\ast}$ states, in contrast to experimental reults (the ``1/2 vs 3/2" puzzle).
One possible  weakness common to these theoretical approaches is that they are derived in the heavy quark limit and
corrections might
be  large. For instance, it is expected that $1/m_c$ corrections  induce a significant mixing
 between $D_1$ amd $D_1^\prime$, which could soften the 1/2-3/2 puzzle at least for the $1^+$
states \cite{Klein:2015doa}. However,  no real conclusion can be drawn until more data
on the masses and the widths of the orbitally excited $D$
meson states become available.
The other puzzle   is that
 the sum of the measured semi-leptonic exclusive rates having $D^{(\ast)}$ in the final state is less than
the inclusive one (``gap" problem) \cite{Liventsev:2007rb, Aubert:2007qw}.
Indeed, decays into $D^{(\ast)}$ make up $\sim$ 70\% of the total inclusive $ B \to X_c l \bar \nu$ rate and decays into $D^{(*)} \pi$  make up another $\sim $ 15\%, leaving a gap of about 15\%.
In 2014 the full BABAR data set has been used
to improve the precision on decays involving $D^{(*)} \pi\, l \, \nu$  and to search for decays of
the type $D^{(*)} \pi\,\pi  l \, \nu$. Preliminary results assign about 0.7\% to  $D^{(*)} \pi\,\pi  l \, \nu$,
reducing the significance of the gap from $7\sigma$ to $3 \sigma$  \cite{Lueck:2015gbr}.

The $B \to D^{\ast\ast} l \nu $  channel has been investigated together with its counterpart with $s$ quark, including the full lepton mass dependence \cite{Bernlochner:2016bci}.
Lattice studies are in progress with realistic  charm mass,
and preliminary results on  $\bar B \to D^{\ast \ast } l \nu$ form factors are available \cite{Atoui:2013sca, Atoui:2013ksa, Blossier:2014kda}.

\subsection{Inclusive  decays}
\label{subsectionInclusive decays}

In  inclusive $ B \rightarrow X_c \, l \, \nu_l$ decays,  the final state
$X_c$ is an hadronic state originated by the charm  quark. There is no dependence on the details of the final state, and quark-hadron duality is generally assumed.
Sufficiently inclusive quantities (typically the width
and the first few moments of kinematic distributions) can be expressed as a double series in $\alpha_s$ and $\Lambda_{QCD}/m$, in the framework of   the Heavy Quark Expansion (HQE),  schematically indicated as
\begin{equation}
\Gamma(B\rightarrow X_c l \nu)=\frac{G_F^2m_b^5}{192 \pi^3}
|V_{cb}|^2 \left[ c_3 \langle O_3 \rangle +
c_5\frac{ \langle O_5 \rangle }{m_b^2}+c_6\frac{ \langle O_6 \rangle }{m_b^3}+O\left(\frac{\Lambda^4_{QCD}}{m_b^4},\; \frac{\Lambda^5_{QCD}}{m_b^3\, m_c^2}, \dots \right)
\right] \label{HQE}
\end{equation}
Here  $c_d$ ($d=3,5,6 \dots$) are short distance coefficients, calculable  in perturbation theory as a series in the strong coupling $\alpha_s$, and
$O_d$ denote local operators of (scale) dimension $d$.
The hadronic
expectation values of the operators  $\langle O_d \rangle $ encode the
nonperturbative corrections and can be parameterized in terms of  HQE  parameters,
whose number grows with
powers of $\Lambda_{QCD}/m_b$.
These parameters are  affected by the
 particular theoretical framework (scheme) that is
used to define the quark masses.
Let us observe that the first order in the series corresponds to the parton order, while  terms of order $\Lambda_{QCD}/m_b$ are absent. At highest orders in $\Lambda_{QCD}/m_b$,   terms including powers of
 $\Lambda_{QCD}/m_c$, sometimes dubbed intrinsic charm  contributions, have to be considered as well \cite{Bigi:2005bh, Breidenbach:2008ua, Bigi:2009ym}.  Indeed, roughly speaking, since $m^2_c \sim O( m_b \Lambda_{\mathrm{QCD}})$ and $\alpha_s(m_c) \sim O(\Lambda_{\mathrm{QCD}})$, contributions of order
 $\Lambda^5_{\mathrm{QCD}}/m^3_b \, m^2_c$
and $\alpha_s(m_c) \Lambda^4_{\mathrm{QCD}}/m^2_b\, m^2_c
$  are expected
comparable in size to  contributions of order $\Lambda^4_{\mathrm{QCD}}/m^4_b$.

At high orders in the HQE the number of new, nonperturbative parameters grows
dramatically. At leading order, the matrix elements can  be reduced to one, while at dimension-four heavy-quark symmetries and the equations of motion
ensure that the forward matrix elements of the operators can be expressed in terms of the
matrix elements of higher dimensional operators. The first nontrivial contributions appear at
dimension five, where two independent parameters are needed, and two independent parameters are also needed at dimension six.
At dimension seven  and eight, nine and  eighteen independent matrix elements appear, respectively, and for higher orders one has an almost
factorial increase of the number of independent parameters.

%%%%%%%%%%%%%%
\begin{table}%[t]
\begin{tabular}{lrr}
\hline
 \hline
  \tablehead{1}{l}{b}{ \color{red}{Exclusive decays}}
& \tablehead{1}{r}{b}{\color{red}{ $ |V_{cb}| \times  10^{3}$}}
  \\
\hline
{\color{blue}{ $\bar{B}\rightarrow D^\ast \, l \, \bar{\nu}$ }}  & \\
\hline
%\hline
FLAG 2016 \cite{Aoki:2016frl} & $ 39.27 \pm 0.49_{\mathrm{exp}} \pm 0.56_{\mathrm{latt}} $ \\
FNAL/MILC 2014 (Lattice $\omega=1$) \cite{Bailey:2014tva}   & $ 39.04 \pm 0.49_{\mathrm{exp}} \pm 0.53_{\mathrm{latt}} \pm 0.19_{\mathrm{QED}} $ \\
%\hline
HFAG 2012 (Sum Rules) \cite{ Gambino:2010bp, Gambino:2012rd, Amhis:2012bh} & $   41.6\pm 0.6_{\mathrm{exp}}\pm 1.9_{\mathrm{th}} $ \\
\hline
{\color{blue}{ $  \bar{B}\rightarrow D \, l \, \bar{\nu} $ }} &   \\
\hline
Global fit   2016 \cite{Bigi:2016mdz}  &  $40.49 \pm 0.97$
\\
Belle 2015 (CLN)   \cite{Glattauer:2015teq,Lattice:2015rga}  & $ 39.86 \pm 1.33 $
 \\
Belle 2015 (BGL)   \cite{Glattauer:2015teq, Lattice:2015rga, Na:2015kha}  & $40.83 \pm 1.13$
 \\
FNAL/MILC  2015 (Lattice  $\omega \neq 1)$ \cite{Lattice:2015rga}  & $39.6 \pm 1.7_{\mathrm{exp+QCD}} \pm 0.2_{\mathrm{QED}} $\\
HPQCD  2015 (Lattice $\omega \neq 1)$  \cite{Na:2015kha}  & $40.2 \pm 1.7_{\mathrm{latt+stat}} \pm 1.3_{\mathrm{syst}}
$
 \\
 \hline
 \tablehead{1}{l}{b}{ \color{red}{Inclusive decays}}
  \\
\hline
 Gambino et al. 2016 \cite{Gambino:2016jkc}  & $42.11 \pm 0.74$ \\
 HFAG  2014  \cite{Amhis:2014hma} & $ 42.46 \pm 0.88 $ \\
\hline
 \tablehead{1}{l}{b}{ \color{red}{Indirect fits}}
\\
\hline
UTfit  2016 \cite{Utfit16} &
$ 41.7 \pm  1.0 $
\\
CKMfitter  2015 ($3 \sigma$) \cite{CKMfitter} &
$41.80^{+0.97}_{-1.64}$
\\
\hline
\hline
\end{tabular}
\caption{Status of exclusive  and  inclusive $|V_{cb}|$  determinations}
\label{phidectab2}
\end{table}
%%%%%%%%%%%%%%%%%%%%%%%%%%%%%%%%%%%%%%

At order $1/m_b^0$ in the HQE, that is the parton level,  the  perturbative corrections up to order $\alpha_s^2$ to the width and to the moments of the lepton energy and hadronic mass
distributions are known completely (see Refs. \cite{Trott:2004xc, Aquila:2005hq, Pak:2008qt, Pak:2008cp, Biswas:2009rb}
and references therein). The terms of order $\alpha_s^{n+1} \beta_0^n$, where $\beta_0$ is the first coefficient of the QCD $\beta$ function, have also been computed following  the
 BLMBrodsky Lepage Mackenzie (BLM) procedure \cite{Aquila:2005hq, Benson:2003kp}.

The next order  is $ \Lambda_{QCD}^2/m_b^2$, and at this order the HQE includes
two  operators, called the kinetic energy  and the chromomagnetic operator.
Perturbative corrections to the coefficients   of the kinetic operator  \cite{Becher:2007tk,Alberti:2012dn}
and  the chromomagnetic operator
\cite{Alberti:2013kxa, Mannel:2014xza, Mannel:2015jka}   have been
 evaluated   at order $\alpha_s^2$ .

Neglecting  perturbative corrections, i.e. working at tree level,  contributions to various observables   have been
computed at order $1/m_b^3$ \cite{Gremm:1996df} and estimated at order $1/m_b^{4,5}$  \cite{Dassinger:2006md, Mannel:2010wj,Heinonen:2014dxa, Heinonen:2016cwm}.

A global fit   is a simultaneous fit to
 HQE  parameters, quark masses and absolute values of  CKM matrix elements obtained by  measuring
 spectra plus all
available moments.
The
HFAG global fit \cite{Amhis:2012bh}  employs as  experimental inputs  the (truncated) moments of the
lepton energy $E_l$  (in the $B$ rest frame) and the $m_X^2$  spectra in $B \to X_c l \nu$ .
It is performed in the kinetic scheme, includes 6  non-perturbative  parameters ($m_{b,c}$, $\mu^2_{\pi,G}$,  $\rho^3_{D,LS}$) and the NNLO $O(\alpha_s)$ corrections,   yielding
  \beq |V_{cb}| = (42.46 \pm 0.88) \times 10^{-3} \label{HFAGincl16}\eeq
In the same kinetic scheme,  another global fit, including the complete power corrections up to $O(\alpha_s\Lambda_{QCD}^2/m_b^2)$, has been performed, giving the estimate $|V_{cb}| = (42.21 \pm 0.78) \times 10^{-3}$ \cite{Alberti:2014yda}. More recently, the effect of including  $1/m_b^{4, 5}$ corrections in the global fit has  been also analyzed, in the so-called Lowest-Lying State Approximation (LLSA), which assumes that the lowest lying heavy meson states
saturate a sum-rule for the insertion of a heavy meson
state sum \cite{Mannel:2010wj, Heinonen:2014dxa, Gambino:2016jkc}. The  LLSA  was used because of the large number of new parameters, in order to provide
loose constraints on the higher power matrix elements.
A resulting  global fit to the semileptonic moments  in  the LLSA gives the estimate  \cite{Gambino:2016jkc}
\beq
|V_{cb}| = (42.11 \pm 0.74) \times 10^{-3}
\label{gambincl16}
\eeq
The  results   \eqref{HFAGincl16}  and \eqref{gambincl16} have practically the same average value, and the uncertainty of about 2\% and 1.8\%, respectively.

%The  HFAG has also performed a  global fit is performed in the 1S scheme, which relies on the calculations of the spectral moments described in Ref \cite{Bauer:2004ve},
%obtaining a similar value
%  \beq |V_{cb}| = (41.98 \pm 0.45) \times 10^{-3}\eeq

Semi-inclusive analyses of  $B_s \to D_s^- X l^+ \nu$ and $B_s \to D_s^{\ast -} X l^+ \nu$ decays and measurements of their branching fractions have been  performed by the \D0   \cite{Abazov:2007wg}
and the LHCb \cite{Aaij:2011ju} experiments, and, more recently, by
Belle \cite{Oswald:2015dma}. Belle has also reported the first measurement of
 the semi-inclusive branching fractions  ${\cal{B}} ( B_s \to D_s X l \nu )$ and
 ${\cal{B}}( B_s \to D^\ast_s X l \nu )$ using its entire $\Upsilon(5S)$ dataset.
The inclusive  semileptonic  branching  fraction  of $ B_s \to X l \nu$
decays was recently measured by Belle and BaBar \cite{Lees:2011ji, Oswald:2012yx}
and found to be in agreement with the expectations from
SU(3)  flavor symmetry \cite{Gronau:2010if, Bigi:2011gf}.

\subsection{$|V_{cb}|$ determinations: recap}

Inclusive and exclusive determinations of $|V_{cb}|$ are collected in Table \ref{phidectab2}. The most precise estimates of $|V_{cb}|$ come from lattice determinations in the $B \to D^\ast$ channel, followed by determinations based on inclusive measurements. They all stay below 2\%  uncertainty. The uncertainty on the determination from $B \to D$ semileptonic decays has recently decreased, and some determination almost reach the  2\% limit from above. The old tension between exclusive and inclusive determinations seems to be confirmed by choosing to compare  the determinations  which claim the most precise estimates. Indeed, by considering the latest inclusive determination \cite{Gambino:2016jkc} and the latest  $B \to D^\ast$ FNAL/MILC  lattice result  \cite{Bailey:2014tva}, the tension amounts to about $3 \sigma$.  The tension is confirmed by the lattice determinations in the
$B \to D$ channel  \cite{Bigi:2016mdz}. It fades by comparing the same  inclusive determination with the (considerable less precise) exclusive   determination based on the sum rule calculation of the $B \to D^\ast$ form factor \cite{Gambino:2010bp, Gambino:2012rd, Amhis:2012bh}.

It is  possible to determine $|V_{cb}|$ indirectly, using
the CKM unitarity relations together with CP violation
and
flavor data, excluding direct informations on decays.
The indirect fit  provided  by the UTfit collaboration \cite{Utfit16} gives
\beq  |V_{cb}|  = (41.7 \pm  1.0) \times 10^{-3} \eeq
while the CKMfitter collaboration (at $3 \sigma$) \cite{CKMfitter} finds
\beq |V_{cb}|  = (41.80^{+0.97}_{-1.64}) \times 10^{-3} \eeq
Indirect fits  prefer a value for $|V_{cb}|$ that is closer to the (higher)
inclusive determination.

\section{Heavy-to-light semileptonic decays}

\subsection{Exclusive  decays}
\label{Exclusivesemi-leptonicdecays33}

%Currently the most precise determination of
%$|V_{ub}|$
%comes from
%charmless semileptonic $B$
%decays, using exclusive or inclusive methods that rely on the measurements of the branching fractions and the %corresponding theoretical inputs.
%For heavy to light  transitions like $B \to \pi l \nu$, $B \to \rho l \nu$,
%and so on,   the impact of  heavy quark symmetry is
%less significant and is mostly reduced to flavour  symmetry relations among $B$
%and $D$ decay semileptonic form factors.

The CKM-suppressed decay $B \to \pi l \nu$ is the most relevant exclusive channel for the determination
of $|V_{ub}|$.
In the
approximation where the leptons are massless,
 the differential rate for  $ B \to \pi l  \nu$ decay reads
\beq
\frac{d \Gamma( B \rightarrow \pi l  \nu)}{dq^2}= \frac{G_F^2  |\overrightarrow{p}_\pi|^3}{24 \pi^3} |V_{ub}|^2 \, |f_+(q^2)|^2
\label{Btopiln}
\eeq
where  $\overrightarrow{p_\pi}$ is the momentum of the pion in the $B$ meson rest frame and $q$ is the momentum of the lepton pair, ranging  $0< q^2< (m_B-m_\pi)^2  \simeq 26.4$ GeV.
The form factor $f_+(q^2)$ refers to the  parameterization of the matrix element between an heavy meson $H$ and  a light pseudoscalar $P$ as
\beq
\langle P(p_P)| V^\mu | H(p_H) \rangle = f_+(q^2) \left (p_H^\mu+p_P^\mu-\frac{m^2_H-m_P^2}{q^2} q^\mu \right) +
 f_0(q^2) \frac{m^2_H-m_P^2}{q^2} q^\mu
 \label{LC1}
\eeq
 Non perturbative  theoretical predictions for form factors are usually confined to particular regions of $q^2$. Complementary regions  are spanned by
LCSR (low $q^2$) and lattice QCD  (high $q^2$).

The process $B \to \pi l \nu$ for light final leptons is well-controlled experimentally; the current
experimental dataset includes various BaBar and Belle measurements, be it untagged or with
different tagging methods \cite{Hokuue:2006nr, Aubert:2006ry, Aubert:2008bf, delAmoSanchez:2010af, Ha:2010rf, Lees:2012vv, Sibidanov:2013rkk}.

%
%
%
%By using current lattice
%QCD methods, the hadronic
%amplitudes for  $\bar B \rightarrow \pi l \bar \nu_l$  can be calculated quite accurately  because  there is only a single %stable
%hadron in both the initial and final states.

The  first lattice determinations of  $f_+(q^2)$  in the $B \to \pi l \nu$ channel, based on unquenched  simulations and obtained by  the HPQCD
\cite{Dalgic:2006dt} and  the Fermilab/MILC \cite{Bailey:2008wp} collaborations, were  in substantial agreement.
In the quenched approximation,  calculations using the  $O(\alpha_s)$ improved Wilson fermions and  $O(\alpha_s)$ improved currents have been performed  on a fine lattice (lattice spacing $a \sim 0.04$ fm) by the QCDSF collaboration \cite{AlHaydari:2009zr} and on a coarser one  (lattice spacing $a \sim 0.07$ fm) by the APE collaboration \cite{Abada:2000ty}.
In 2015, the Fermilab/MILC collaboration has  presented its latest $|V_{ub}|$ determination,
based on the MILC asqtad 2+1-flavor lattice configurations at four different lattice spacings \cite{Lattice:2015tia}.  Light-quark masses have been set down to 1/20 of the physical strange-quark
mass and the limit to the continuum has been obtained using staggered chiral perturbation
theory in the hard-pion and SU(2) limits. The form factors have been extrapolated from large-recoil momentum (17 GeV$^2 < q^2< $26 GeV$^2$) to the full kinematic range by means of a
parameterization based on the BCL $z$-expansion.
By combining the lattice form factors with recent experimental data from  Babar  and Belle Collaborations,
they obtain  \cite{Lattice:2015tia}
\beq |V_{ub}|  = (3.72 \pm  0.16) \times 10^{-3}\eeq
where the error reflects both the lattice and experimental uncertainties, which are now on par with
each other.

In 2015, a new determination has also been presented    by the RBC/UKQCD  collaboration \cite{Flynn:2015mha},
yielding, at $N_f=  2 + 1$
\beq |V_{ub}|  = (3.61 \pm 0.32) \times 10^{-3}\eeq

Preliminary results on form factors   by the  ALPHA \cite{Bahr:2012vt, Bahr:2012qs} and  HPQCD \cite{Bouchard:2012tb}
($N_f=  2 + 1$) collaborations, were updated and published in 2016.
The HPQCD collaboration has presented 2+1+1-flavor results \cite{Colquhoun:2015mfa}; they have given the first lattice QCD calculations of
$ B \to \pi l  \nu$ decay for $u/d$ quark
masses going down to their physical values, calculating the $f_0$ form factor at zero recoil to 3\% precision.
The
ALPHA collaboration has presented a  2-flavor calculation  of semi-leptonic form factors for $B_s$-mesons  \cite{Bahr:2016ayy}, but the challenges faced have implications also for $B$ meson decays. For the first time, they were able to perform a study of the continuum limit of fully
non-perturbatively renormalized form factors.

%

%%%%%%%%%%%%%%%%%%%%%%%%%%%%%%%%%%%%%%%%%%%%%%
In 2016 the FLAG working group has proposed the $|V_{ub}|$
 estimate ($N_f=2+1$)  \cite{Aoki:2016frl}
\beq |V_{ub}|  = (3.62 \pm 0.14) \times 10^{-3}\eeq
As already mentioned, the experimental value of $|V_{ub}||f_{+}(q_2)|$
can be extracted from the measured branching fractions for $B_0 \to \pi^+ l \nu$  and/or $B^{\pm} \to \pi_0  l\nu$; then
$|V_{ub}|$ can  be determined by performing fits to the
parameterization of the form factor $|f_{+}(q_2)|$.
The parameterization adopted by FLAG is the
constrained BCL
$z$ parameterization. The fit can be done in two
ways: one option is to perform separate fits to lattice and experimental results, and extract
the value of $|V_{ub}|$ from the ratio of the respective  coefficients; a second option is to perform
a simultaneous fit to lattice and experimental data, leaving their relative normalization
$|V_{ub}|$ as a free parameter. FLAG adopts the second strategy, leading to a smaller uncertainty on $|V_{ub}|$\cite{Aoki:2016frl}.

HFAG has performed a simultaneous fit to  the four most precise measurements from BABAR and
Belle
and the 2009 Fermilab/MILC lattice calculations \cite{Bailey:2008wp},  yielding \cite{Amhis:2014hma}
\beq |V_{ub}|  = (3.28 \pm  0.29) \times 10^{-3}\eeq

At
large recoil (small $q^2$),
 direct LCSR calculations of the semi-leptonic  form factors  are available, which have benefited by   progress in pion distribution amplitudes, next-to-leading and leading  higher order twists and QCD corrections (see e.g. Refs.~\cite{Khodjamirian:2011ub, Bharucha:2012wy,Li:2012gr,Wang:2015vgv, Shen:2016hyv} and references within).
The latest HFAG
simultaneous fit uses a recent result for $f_+(0)$ from
LCSR \cite{Bharucha:2012wy}, and gives
\beq |V_{ub}|  = (3.53 \pm  0.29) \times 10^{-3}\eeq

Another LCSR determination  of $|V_{ub}| $ uses a Bayesian  uncertainty analysis of the
$B \to \pi$
 form factor and combined BaBar/Belle data to obtain \cite{Imsong:2014oqa}
\beq
|V_{ub}|= (3.32^{+0.26}_{-0.22})\, \times\,  10^{-3}
\eeq

Belle  has recently found  a branching ratio of
$ {\cal{B}} ( B^0 \to \pi^- l^+ \nu) = (1.49 \pm 0.09_{\mathrm{stat}} \pm 0.07_{syst}) \times 10^{-4}$ \cite{Sibidanov:2013rkk},
which is competitive with the more precise
results from untagged measurements.
By employing
this measured partial branching fraction, and combining    LCSR, lattice points and the BCL \cite{Bourrely:2008za}
 description of the
$f_+(q^2)$ hadronic form factor,
 Belle extracts the value
\beq |V_{ub}|  = (3.52 \pm  0.29) \times 10^{-3}\eeq
All the above mentioned exclusive determinations
 have been reported  in Table \ref{phidectab03}.
 %, together with inclusive ones and indirect fits.
Other values for exclusive $|V_{ub}|$ have been computed  in the relativistic quark model \cite{Faustov:2014zva}.
In 2016, Belle has  also presented the first experimental result on $B \to \pi \tau \nu$, with an upper limit compatible with the SM \cite{Belle16}.

%%%%%%%%%%%%%%
\begin{table}%[t]
\begin{tabular}{lrr}
\hline
  \tablehead{1}{l}{b}{ \color{red}{Exclusive decays}}
& \tablehead{1}{r}{b}{\color{red}{ $ |V_{ub}| \times  10^{3}$}}
  \\
\hline
{\color{blue}{ $\bar B \rightarrow \pi l \bar \nu_l$   }}    & \\
\hline
FLAG 2016    \cite{Aoki:2016frl}  & $3.62 \pm 0.14$\\
Fermilab/MILC  2015  \cite{Lattice:2015tia}  & $3.72 \pm 0.16$\\
RBC/UKQCD   2015  \cite{Flynn:2015mha}  & $3.61 \pm 0.32$\\
HFAG 2014 (lattice)    \cite{Amhis:2014hma}  & $3.28 \pm 0.29$\\
HFAG 2014 (LCSR)  \cite{Khodjamirian:2011ub, Amhis:2014hma}  & $3.53 \pm 0.29$   \\
Imsong et al. 2014 (LCSR, Bayes an.)     \cite{Imsong:2014oqa}  & $3.32^{+0.26}_{-0.22}$\\
Belle 2013 (lattice + LCSR)     \cite{Sibidanov:2013rkk}   & $3.52 \pm 0.29$ \\
\hline
{\color{blue}{ $\bar B \rightarrow \omega l \bar \nu_l$  }}     & \\
\hline
Bharucha et al. 2015 (LCSR)     \cite{Straub:2015ica}  & $3.31 \pm 0.19_{\mathrm{exp}} \pm 0.30_{\mathrm{th}}$\\
\hline
{\color{blue}{ $\bar B \rightarrow \rho l \bar \nu_l$   }}    & \\
\hline
Bharucha et al. 2015 (LCSR)     \cite{Straub:2015ica}   & $3.29 \pm 0.09_{\mathrm{exp}} \pm 0.20_{\mathrm{th}}$\\
\hline
{\color{blue}{ $ \Lambda_b \rightarrow p \, \mu\nu_\mu$   }}    & \\
\hline
LHCb  (PDG)    \cite{Fiore:2015cmx}    & $ 3.27  \pm 0.23 $\\
\hline
 \tablehead{1}{l}{b}{ \color{red}{Indirect fits}}
\\
\hline
UTfit  (2016) \cite{Utfit16} &
$3.74 \pm  0.21$\\
CKMfitter  (2015, $3 \sigma$) \cite{CKMfitter} &
$ 3.71^{+0.17}_{-0.20}$
\\
\hline
\end{tabular}
\caption{Status of  exclusive $|V_{ub}|$  determinations and indirect fits}
\label{phidectab03}
\end{table}
%%%%%%%%%%%%%%

Recently, significantly improved branching ratios of
 heavy-to-light semi-leptonic decays other than $B \to \pi$  have been reported, that reflect on
increased precision
for $|V_{ub}|$ values inferred by these decays.
In 2010,
the Babar collaboration has started  investigating  the  $ B \to \rho l \nu$ channel  \cite{delAmoSanchez:2010af}.
By comparing the
measured distribution in $q^2$ (with an upper limit at $q^2 = 16$ GeV)  and using
the 2004 LCSR  predictions for the form factors \cite{Ball:2004rg},
the estimate
$
|V_{ub}|  =
(2.75 \pm 0.24 ) \times 10^{-3}  $ is given, and compared with  the $
|V_{ub}|  =
(2.83 \pm 0.24 ) \times 10^{-3}  $
 estimate obtained using
 the old ISGW2 quark model \cite{ Scora:1995ty}. Both values are
lower than the ones extracted by $B \to \pi$ decays.
More recent results have been provided by the Belle Collaboration \cite{Sibidanov:2013rkk}.

Measurements of $ B \to \omega l \nu$ decays have been reported by Belle \cite{Schwanda:2004fa, Lees:2013gja, Sibidanov:2013rkk} and Babar \cite{Lees:2012vv, Lees:2012mq}.
In the Belle tagged analysis  \cite{Sibidanov:2013rkk}, which includes results for several  exclusive channels, an evidence of a broad resonance around 1.3
GeV dominated by the
$B^+\to f_2 l \nu$
decay has also been reported for the first time.
In the Babar analysis with semileptonically tagged B mesons  \cite{Lees:2013gja}, the value of
$|V_{ub}|$ has been extracted, yielding
$
|V_{ub}|  =
(3.41\pm 0.31 ) \times 10^{-3}  $, using the LCSR form factor determination \cite{Ball:2004rg}, and
$
|V_{ub}|  =
(3.43\pm 0.31 ) \times 10^{-3} $
using the ISGW2 quark model \cite{Scora:1995ty}.
A major problem is that the quoted uncertainty does not include any uncertainty from theory, since uncertainty estimates of the
form-factor integrals are not available.

In 2015, new LCSR computations have been performed for the form factors of $B \to \rho$, $B \to \omega$, $B_s \to K^\ast$
and $B_s \to \phi$ semileptonic decays from  light-cone  sum  rules  using  updated  hadronic  input  parameters \cite{Straub:2015ica}.
By combining the form factor computation for $ B \to \omega l \nu$ decays at $q^2 < 12 $ GeV  with  2012 Babar data \cite{ Lees:2012vv,Lees:2012mq}, they  found
\beq
|V_{ub}|  =
(3.31 \pm 0.19_{\mathrm{exp}} \pm 0.30_{\mathrm{th}}) \times 10^{-3}  \eeq
and combining the form factor computation for $ B \to \rho l \nu$ decays at the same $q^2 < 12 $ GeV with  Belle data \cite{Sibidanov:2013rkk}, they found
\beq
|V_{ub}|  =
(3.29 \pm 0.09_{\mathrm{exp}} \pm 0.20_{\mathrm{th}})  \times 10^{-3}  \eeq

The branching fractions for
 $ B \rightarrow \eta^{(\prime)} l \nu $ decays have been  measured in 2007 by the CLEO Collaboration \cite{Adam:2007pv} and, more recently, by the BaBar collaboration
 \cite{delAmoSanchez:2010zd}.
The
value of the ratio
\beq \frac{{\cal{B} }(B^+ \to \eta^\prime l^+ \nu_l)}{{\cal{B} }(B^+ \to \eta l^+ \nu_l)}=0.67 \pm 0.24_{\mathrm{stat}} \pm 0.11_{\mathrm{syst}} \eeq
seems to
allow an important gluonic singlet contribution to the
$\eta^\prime$
form factor \cite{delAmoSanchez:2010zd, DiDonato:2011kr}. In 2015 the
analysis of all $B$, $B_s$  $\to \eta^{(\prime)}$ form factors  has been performed
in the LCSR framework \cite{Duplancic:2015zna}. At Belle, the branching ratio
${\cal{B} }(B \to \eta l \nu_l)= (0.42\pm 0.11_{\mathrm{stat}} \pm 0.03_{\mathrm{syst}}) \times 10^{-4}$ has been measured, and an upper limit at 90\% to the branching ratio for $B \to \eta^\prime l \nu_l $ set of $0.76 \times 10^{-4}$ \cite{Belle16}.

Another channel that proceeds at the lowest partonic level in the SM from the $ b \to u l \nu$ decay  is the $B \to \pi \pi l \nu$ one, that gives the possibility to explore several angular observables, of extracting $|V_{ub}|$  and involves  resonant contributions of vector and scalar mesons \cite{Faller:2013dwa, Kang:2013jaa}.

The $B_s \to K^{(\ast) }l \nu$ decays have not been measured yet, though, being expected to be within the reach of future $B$-physics results, can become an additional channel to extract $|V_{ub}|$  \cite{Flynn:2015mha, Meissner:2013pba, Albertus:2014rna, Bouchard:2014ypa, Feldmann:2015xsa}.

Although this report focusses on meson deays, it is interesting to compare with another decay  depending on  $|V_{ub}|$, that is the
baryonic  semileptonic $\Lambda^0_b \to p l^- \bar \nu$ decays, where,  at the parton level, the $b$-quark decays  into the $u$-quark emitting a $W^-$ boson \cite{Stone:2014mza, Khodjamirian:2011jp, Detmold:2013nia}.
In the  Run I of the LHC (2011-2012), the LHCb experiment
collected an integrated luminosity of 3 fb$^{-1}$ at center of mass energies of $\sqrt{s}=7$ TeV and
$\sqrt{s}=8$ TeV. At the end of Run I,
LHCb measured the ratio of decay rates for $\Lambda^0_b \to p l^- \bar \nu$, at high $q^2$, where the background is expected to be reduced and lattice predictions more reliable.
This result has been combined with the ratio of form factors computed
using lattice QCD with
2+1
flavors of dynamical domain-wall fermions \cite{Detmold:2015aaa},
 enabling the first determination of the ratio of CKM elements $|V_{ub}|/|V_{cb}|$  from
baryonic decays \cite{Aaij:2015bfa}. The channel $\Lambda_b \to \Lambda_c (\to p K \pi) \mu \bar \nu$ has been used as a control channel.
To compute the form factors  $\Lambda_b \to p$  and $\Lambda_b \to \Lambda_c$, the $b$
and $c$-quarks have been implemented with relativistic
heavy-quark actions and  the lattice
computation  has been  performed  for  six  different  pion  masses  and  two  different  lattice  spacings,  using
gauge-field configurations generated by the RBC and UKQCD collaborations; the form factor
results were extrapolated to the physical pion mass and the continuum limit, parameterizing the
$q^2$-dependence using $z$ expansions \cite{Detmold:2015aaa}.
The 2015 LHCb result reads \cite{Aaij:2015bfa}
\beq
\frac{|V_{ub}|}{|V_{cb}|}= 0.083 \pm 0.004_{\mathrm{exp}} \pm 0.004_{\mathrm{latt}}
\label{ratiovcbcub}
\eeq
The value of $|V_{ub}|$ depends on the choice of the value of $|V_{cb}|$. By taking the inclusive determination $|V_{cb}|_{incl}= (42.21 \pm 0.78 )\times 10^{-3}$, the value
$|V_{ub}|= (3.50 \pm 0.17_{\mathrm{exp}} \pm 0.17_{\mathrm{FF}} \pm 0.06_{\mathrm{|V_{cb}|}} )\times 10^{-3}$ is obtained \cite{Rosner:2015wva}, where the errors are
from experiment, the form factors, and
$|V_{cb}|$, respectively.
This result is 1.4$\sigma$ lower than
 the determination from leptonic decays.
By taking instead the higher value of the exclusive determination $|V_{cb}|= (39.5 \pm 0.8) \times 10^{-3}$, given by PDG 2014 \cite{Agashe:2014kda}, the LHCb reports \cite{Fiore:2015cmx}
\beq
|V_{ub}|= (3.27 \pm 0.23) \times 10^{-3}
\eeq
Let us observe the value of  $|V_{ub}|/|V_{cb}|$ extracted, from the same LHCb data, by using  the  relativistic quark model, is about 1.4 higher than Eq. (\ref{ratiovcbcub})\cite{Faustov:2016pal}. By  using $|V_{cb}|_{incl}$, it translates into an higher value for $|V_{ub}|$, in better agreement with the inclusive determination rather than the exclusive one favoured by lattice results.

\subsection{Inclusive   decays}
\label{vuninclusivo}

The extraction of $|V_{ub}|$ from inclusive decays requires to address theoretical issues absent in the inclusive $|V_{cb}|$ determination.
OPE techniques  are not applicable in the   so-called  endpoint or singularity or  threshold phase space region,
 corresponding to the kinematic region near the limits of
both the lepton energy  $E_l$ and $q^2$ phase space, where the rate is dominated by
the production of low mass final hadronic states.
This region  is sensitive to the Fermi motion of the $b$
quark inside the $B$
meson. It is also
plagued by the presence
 of large double (Sudakov-like)  perturbative  logarithms at all orders in the strong coupling.
Corrections  can be large  and need to be resummed at all orders \footnote{See e.g. Refs.~ \cite{DiGiustino:2011jn,Aglietti:2007bp,Aglietti:2005eq,Aglietti:2005bm,Aglietti:2005mb,Aglietti:2002ew, Aglietti:2000te, Aglietti:1999ur} and references therein.}. The kinematics cuts due to the large $B \to X_c l \nu$ background enhance the weight of the threshold region with respect to the case of
  $b \rightarrow c$ semi-leptonic decays; moreover, in the latter, corrections are not expected  as singular as in the $ b \rightarrow u$ case, being  cutoff by the charm mass.

On the experimental side, efforts have been made
 to control the background and access to a large part of the phase space, so as to reduce,
on the whole,  the weight of the endpoint region.
 Latest results by Belle \cite{Urquijo:2009tp} and BaBar \cite{Lees:2011fv}
use their complete data sample, $ 657$ x $ 10^{6}$  $B$-$\xbar{B}$ pairs for Belle   and 467 x $ 10^{6}$ $B$-$\xbar{B}$ pairs for BaBar. Although the two analyses differ
in the treatment of the background, both collaborations claim to access $\sim  90$\% of the phase space.

On the theoretical side, several schemes are available. They assume an underlying $b$-quark decaying, since
weak annihilation contributions seems to be strongly constrained by semileptonic charm decays \cite{Gambino:2010jz,Bigi:2009ym,Ligeti:2010vd}.
All of the schemes are  tailored
to analyze data in the threshold region,  but  differ significantly
in their treatment of perturbative corrections and the
parametrization of non-perturbative effects.

The analyses from BaBar \cite{Lees:2011fv}  and Belle \cite{Urquijo:2009tp}  collaborations, as well as  the latest HFAG averages \cite{Amhis:2014hma},
rely on at least four theoretical different QCD calculations of the inclusive partial
decay rate: ADFR by Aglietti, Di Lodovico, Ferrera and Ricciardi \cite{Aglietti:2004fz, Aglietti:2006yb,  Aglietti:2007ik}; BLNP
by Bosch, Lange, Neubert and Paz \cite{Lange:2005yw, Bosch:2004th, Bosch:2004cb}; DGE, the dressed gluon exponentiation, by Andersen and Gardi \cite{Andersen:2005mj}; GGOU by Gambino, Giordano, Ossola and Uraltsev \cite{Gambino:2007rp} \footnote{Recently, artificial neural networks have been used to parameterize the shape functions and  extract $|V_{ub}|$ in the GGOU framework \cite{Gambino:2016fdy}. The results are in good agreement with the original paper.}.
They can be roughly  divided into approaches based on the estimation of the shape function (BLNP, GGOU) and on resummed perturbative QCD (ADFR, DGE).
Although conceptually quite different, all the above approaches generally
lead to roughly consistent results when the same inputs are used and the
theoretical errors are taken into account.
The HFAG estimates \cite{Amhis:2014hma}, together with the latest estimates by BaBar \cite{Lees:2011fv, Beleno:2013jla} and Belle
\cite{Urquijo:2009tp}, are reported in Table \ref{phidectab04}.
\begin{table}%[t]
\begin{tabular}{lrrrr}
 \hline
  \tablehead{1}{l}{b}{ \color{red}{Inclusive decays}
 ($  |V_{ub}| \times  10^{3}$)}
  \\
\hline
&  \color{blue}{ADFR   } \cite{Aglietti:2004fz, Aglietti:2006yb,  Aglietti:2007ik}  & \color{blue}{BNLP   } \cite{Lange:2005yw, Bosch:2004th, Bosch:2004cb}&  \color{blue}{DGE }   \cite{Andersen:2005mj} &   \color{blue}{GGOU  }   \cite{Gambino:2007rp}  \\
\hline
HFAG 2014 \cite{Amhis:2014hma} & $4.05 \pm 0.13^{+ 0.18}_{- 0.11}$ & $ 4.45 \pm 0.16^{+0.21}_{-0.22}  $  & $4.52 \pm 0.16^{+ 0.15}_{- 0.16}$ &
$4.51 \pm  0.16^{ + 0.12}_ { -0.15} $  \\
BaBar 2011  \cite{Lees:2011fv} &  $4.29 \pm 0.24^{+0.18}_{-0.19}  $  & $4.28 \pm 0.24^{+0.18}_{-0.20}  $    & $4.40 \pm 0.24^{+0.12}_{-0.13}  $
& $4.35 \pm 0.24^{+0.09}_{-0.10}  $ \\
 Belle 2009 \cite{Urquijo:2009tp} & $4.48 \pm 0.30^{+0.19}_{-0.19}  $ & $ 4.47 \pm 0.27^{+0.19}_{-0.21}  $ &  $4.60 \pm 0.27^{+0.11}_{-0.13}  $ & $4.54 \pm 0.27^{+0.10}_{-0.11}  $ \\
\hline
\end{tabular}
\caption{Status of inclusive $|V_{ub}|$  determinations}
\label{phidectab04}
\end{table}
The BaBar and Belle  estimates  in Table \ref{phidectab04} refers to the value extracted by
the
most inclusive measurement, namely the one based on
the two-dimensional fit of the $M_X-q^2$
distribution with
no phase space restrictions, except for
$p^\ast_l > 1.0$  GeV. This selection  allow to access approximately
90\% of the total phase space \cite{Beleno:2013jla}.
The BaBar collaboration also
reports measurements of $|V_{ub}|$
in other regions of the phase space \cite{Lees:2011fv}, but the values reported in  Table \ref{phidectab04} are the most precise.
When averaged, the ADFR value is lower than the one obtained with the other three approaches, and closer to the exclusive values;  this difference
disappears
if we restrict to the BaBar
 and Belle results quoted in  Table \ref{phidectab04}.
By taking the arithmetic average of the
results obtained from these  four different QCD predictions of the partial rate the Babar collaboration gives \cite{Lees:2011fv}
\beq
|V_{ub}|=(4.33 \pm 0.24_{\mathrm{exp}} \pm 0.15_{\mathrm{th}}) \times 10^{-3}
\label{VinclBabar}
\eeq
Another HFAG average value is $|V_{ub}|=  4.62 \pm 0.20 \pm 0.29$  \cite{Amhis:2014hma},   obtained from a global fit in the 1S scheme using the BLL (Bauer, Ligeti and Luke) \cite{Bauer:2001rc},
theoretical approach, which is limited to  measurements that use $m_X- q^2$ cut.

\subsection{$|V_{ub}|$ determination: recap}

The parameter $|V_{ub}|$ is the less precisely known among the modules of the CKM matrix elements.
The error on the inclusive determinations is around 4-5\% in the latest averaged HFAG values, in several theoretical schemes.  The uncertainties of the lattice based exclusive determinations of  $|V_{ub}|$, in the $ B \to \pi$ channel, have halved in the last two years, ranging now around 4\%.

By comparing the value (\ref{VinclBabar}) (or results in Table \ref{phidectab04}) with results in Table \ref{phidectab03}, we observe a tension between exclusive and inclusive determinations, of the order of $2-3\sigma$, according to the chosen values.
A lot of theoretical effort has been devoted in the past to clarify the present tension by inclusion of  NP effects, but this possibility is strongly limited
 by electroweak constraints on the effective $Z \bar b b $ vertex \cite{Crivellin:2014zpa}.
%(aggiungi articolo di Bigi 2015)
The difficult interpretation of this tension is enhanced by the fact that experimental access to such a large portion of phase space as 90\%,  once accepted that the difficult background subtraction is trustable, seems to reduce the weight of the shape function region in the inclusive determination, and justify a theoretical description analogous to the one used for the inclusive $B \to X_c$ semileptonic decay, in terms of local OPE.

Within Table \ref{phidectab03}, the values obtained for $B \to \rho/\omega \, l \nu$ appear to be systematically lower than the ones for  $B \to \pi l \nu$. Notwithstanding the large errors, several theoretical possibilities have  been already put forward, as e.g. corrections for the $\rho$ lineshape in $B \to \pi \pi l \nu$ \cite{Kang:2013jaa}. A possibile S-wave  background not separated in  the experimental analysis of $B \to \rho l \nu$ has been  estimated to give a modest contribution \cite{Meissner:2013pba}, and also the effect of right-handed currents \cite{Bernlochner:2014ova, Bernlochner:2016ogh} seems to be disfavoured by data.

It is  possible to determine $|V_{ub}|$ indirectly, using
the CKM unitarity relations together with CP violation
and
flavor data, as done for $|V_{cb}|$.
The indirect fit  provided  by the UTfit collaboration \cite{Utfit16} gives
\beq  |V_{ub}|  = (3.74 \pm 0.21) \times 10^{-3}\eeq
while the CKMfitter collaboration (at $3 \sigma$) \cite{CKMfitter} finds
\beq  |V_{ub}|  = (3.71^{+0.27}_{-0.20}) \times 10^{-3}\eeq
At variance
with the $|V_{cb}|$ case, the results of the global fit prefer a value for $|V_{ub}|$ that is closer to the (lower)
exclusive  determination.
Belle II
is expected, at about 50 ab$^{-1}$,  to decrease experimental  errors on both inclusive and exclusive $|V_{ub}|$  determinations up to about 2\%.

%%%%%%%%%%%%%%

%%%%%%%%%%%%%%

%\section{Rare decays}

%TO UPDATE

%\section{$\tau$ leptons in the final state}

\section{Leptonic decays}

In the SM, the weak decay $B^+ \to l^+ {\nu}_l$ of a charged $B$ meson occurs, in the parton model at the lowest perturbative order, through the annihilation of the heavy and light quark inside the meson, and it is therefore mediated by a charged current.
%
%the  purely leptonic decay  $B \to l {\nu}_l$
The branching ratio is given by
\beq {\cal{B}}(B^+\to l^+  {\nu}_l) = \frac{G_F^2 m_B m_l^2}{8 \pi} \left(1- \frac{m_l^2}{m_B^2} \right)^2 f_B^2 |V_{ub}|^2 \tau_B
\label{lepto1}
\eeq
The decay constant $f_B$ parameterizes the matrix elements of the axial vector current
$
<0|\bar{b} \gamma^\mu \gamma_5 q|B^+(p_B)> = p_B^\mu f_B
$
%
%.
and is calculated on lattice.
For heavy-light mesons, it is sometimes convenient to define and study the quantity
$
\Phi_{B} \equiv f_B \sqrt{m_B}
$
which approaches a constant (up to logarithmic corrections) in the $m_b \to \infty$ limit.
All the other inputs in Eq. (\ref{lepto1}) are
measured experimentally.
 The  branching  fraction  depends strongly on the mass of the lepton due to helicity
suppression,  and  thus the
 $B^+ \to \tau^+  {\nu}_\tau$ decay
is  expected  to  have the largest leptonic branching fraction of the
$B^+$
meson
and is the only decay of this kind for which there is experimental  evidence.
%The only charged current $B$ meson decay  that has been observed so far is the
% $B \to \tau  {\nu}_\tau$ decay,
It was observed for the first time  by  Belle in 2006 \cite{Ikado:2006un}.

$B$
factories  have  studied  purely  leptonic    decays  with  a  goal  of
searching  for  new  physics  beyond  the  SM.
 The  analysis  relies  on  reconstructing  a  hadronic
or  semi-leptonic $B$ decay  tag,   finding  a
candidate  in  the  remaining  track  and  photon
candidates, and examining the extra energy in the event which should be almost zero when a
real $\tau^-$ decays into electron and muons.
The  branching  fraction  measured by Babar using
 semileptonic tags \cite{Aubert:2009wt} and hadronic tags \cite{Lees:2012ju} averaged to \cite{Lees:2012ju}
 \beq
{\cal{B}}( B^+ \to \tau^+ \nu_\tau) =( 1.79 \pm 0.48) \times 10^{-4}
\eeq
in agreement with oldest Belle result using hadronic \cite{Ikado:2006un} and semileptonic tags \cite{Hara:2010dk}.
The results were more than 2$\sigma$ higher than the SM estimate
based on a global fit \cite{CKMfitter}.
However, Belle has recently reanalyzed both samples of their data, using the hadronic \cite{Adachi:2012mm} and semileptonic tags \cite{Kronenbitter:2015kls}. A much lower averaged  branching fraction is given \cite{Kronenbitter:2015kls}
 \beq
{\cal{B}}( B^+ \to \tau^+ \nu_\tau) =( 0.91 \pm 0.22) \times 10^{-4}
\eeq
which is now aligned with the SM predictions, even though the error is about 20\%.
In all measurements there  are  large  statistical and dominant systematical errors, and the significances  are less than 5$\sigma$.
% not  large:  Belle  quotes  4.6$\sigma$
%for  their
%combined hadronic and semileptonic tags, while BaBar quotes 3.3$\sigma$
%and 2.3 $\sigma$ for hadronic
%and semileptonic tags.
 An higher level of  precision  is required to explore the possibility of new physics effects.

The measurement of the  branching fraction  provides a direct experimental determination of the
product  $f_B  |V_{ub}|$, which can be used to determine
$ |V_{ub}|$
when combined with lattice predictions of $f_B$.
Combining the experimental values with the
 mean $B^+$-meson lifetime $\tau_B= 1.641 \pm 0.008$  \cite{Beringer:1900zz}
and their
averages for the $B$ meson decay constant,
%$f_B=190.5 \pm 4.2$ MeV ($N_f=2+1$),
the FLAG  working group obtains a range of values varying with the experimental collaboration and $N_f$. By assuming $N_f= (2,2+1,2+1+1)$, they give \cite{Aoki:2016frl}
\bea
 |V_{ub}|  &=& (3.83 \pm  0.48 \pm 0.15, 3.75 \pm  0.47 \pm 0.09, 3.87 \pm  0.48 \pm 0.09) \times 10^{-3} \quad \quad \mathrm{Belle} \nonumber \\
  |V_{ub}|  &=& (5.37 \pm  0.74 \pm 0.21, 5.26 \pm  0.73 \pm 0.12, 5.43 \pm  0.75 \pm 0.12) \times 10^{-3} \quad \quad \mathrm{Babar}
  \eea
where the first error comes from experiment and the second comes from the uncertainty
in $f_B$.
Another  recent determination  employs averages from Babar and Belle leptonic modes and yields \cite{Rosner:2015wva}
\beq |V_{ub}|  = (4.12 \pm  0.37 \pm 0.06) \times 10^{-3}\eeq
The   $|V_{ub}|$ values  seem to point towards the higher semileptonic inclusive $|V_{ub}|$ determinations, but there is also consistence with the exclusive values, within the large errors.
The accuracy is not yet enough to draw definite conclusions and to make the leptonic  channel competitive for $
|V_{ub}|$ extraction.
At Belle II, the total experimental precision is expected to reduce uncertainty below 5\%  at 50ab$^{-1}$ \cite{BelleII16}.

The  most  stringent  upper  limits for the charged $B$ decaying into muons  have been measured by Babar  \cite{Aubert:2009ar}
${\cal{B}}( B^+ \to \mu^+ \nu_\mu) < 1.0  \times 10^{-6}$, and  for decays into electrons by Belle \cite{Satoyama:2006xn},
${\cal{B}}( B^+ \to e^+ \nu_e) < 9.8  \times 10^{-7}$,  both with untagged methods.

%Finally, let us just mention that search of possible lepton flavour violations can also be made independently of $|V_{ub}|$
%by building ratios of branching fractions, such as
%$
%R^\prime = \tau_{B^0}/\tau_{B^+}\, {\cal{B}}( B^+\to \tau^+ \nu_\tau) /{\cal{B}}( B^0 \to \pi^- l^+ \nu_l)
%$.

\section{Conclusions}

%\begin{comment}
We have summarized the current status of semileptonic and leptonic $B$ decays, which presents significant theoretical and experimental progress. Higher   precision has been attained in challenging measurements, as for instance the branching fractions of   exclusive $B \to \rho/\omega \, l \, \bar \nu$ decays or  fully leptonic charged $B$ decays.
More  results are  expected, at present  from  further analyses of data provided by  the $B$ factories and LHCb, and in the (approaching)  future from Belle II.

On the theoretical side, the perturbative calculations, in general, have reached a phase of maturity, and the larger theoretical errors are due to non-perturbative approaches.
Errors have been recently lowered in  both lattice and LCSR frameworks; new global fits for inclusive processes also sport
 further reduced theoretical uncertainties.
New physics is always more constrained.
Still awaiting firmly established solutions are  a few dissonances within the SM, %such as the so-called ``1/2 vs 3/2"  and ``gap" puzzles, the possibility of flavour violation observed in decays into tauons, and the
 as the long-standing tension between the inclusive and exclusive determination of $|V_{cb}|$  and  $|V_{ub}|$.

 The most precise estimates of $|V_{cb}|$ come from lattice determinations in the $B \to D^\ast$ channel \cite{Bailey:2014tva}, which is  below 2\%  uncertainty; by comparing with the latest inclusive determination \cite{Gambino:2016jkc}, which has almost the same precision,  the tension amounts to about $3 \sigma$. This tension is confirmed by comparing with  determinations from $B \to D$ semileptonic decays, whose  uncertainty  has decreased in some recent computations, bordering the  2\% limit.

The parameter $|V_{ub}|$ is the less precisely known among the modules of the CKM matrix elements.
The error on the inclusive determinations is around 4-5\% in the latest averaged HFAG values, in several theoretical schemes.  The uncertainties of the lattice based exclusive determinations of  $|V_{ub}|$, in the $ B \to \pi$ channel, have halved in the last two years, ranging now around 4\%. The inclusive/exclusive determination tension amounts to about 2-3 $\sigma$.
 %

%\end{comment}

Another  tension with the SM, recently recognized, is the apparent excess of ${\cal{R}}(D^{(\ast)})$
on the SM predictions, which is causing an intense theoretical work to explore the possibility of lepton flavour violation or lepton non-universality.

In contrast to the above, a previous tension within the SM seems to have faded after the
Belle recent reanalysis of
$
{\cal{B}}( B^+ \to \tau^+ \nu_\tau) $,
using both the hadronic \cite{Adachi:2012mm} and semileptonic tags \cite{Kronenbitter:2015kls}.
The results yield an averaged  branching fraction  for the
$
 B^+ \to \tau^+ \nu_\tau
$ channel
much lower than in the past, and the branching fraction
 is now aligned with the SM predictions, even though the error is about 20\%.
 An higher level of  precision  is required to explore the possibility of new physics effects.

Belle II, which has completed the accelerator commissioning phase and should nominally start in fall 2018, is expected to give a substantial boost to understand tensions within the SM. At the ultimate goal of 50 ab$^{-1}$, precision on fully leptonic charged $B$ decays  is foreseen to be reduced below 5\%, on ${\cal{R}}(D^{(\ast)})$ to about 2-3\%, and more than halved for $|V_{ub}|$  inclusive  and exclusive determinations from the $B \to \omega$ channel \cite{Finocchiarotalk,BelleII16}.

%%%%%%%%%%%%%%%%%%%%%%%%%%%%%%%%%%%%%%%%%%%%%%%%
%% BACKMATTER
%%%%%%%%%%%%%%%%%%%%%%%%%%%%%%%%%%%%%%%%%%%%%%%%

%\begin{theacknowledgments}
%GR acknowledges partial support   by
%Italian MIUR under project 2010YJ2NYW and INFN
%under specific initiative QNP.
%\end{theacknowledgments}

\bibliographystyle{aipproc}   % if natbib is available

\bibliography{VxbRef}

\end{document}